\documentclass[pra,twocolumn,showpacs]{revtex4}
\usepackage{graphicx}
\newcommand{\beq}{\begin{equation}}
\newcommand{\eeq}{\end{equation}}
\newcommand{\beqa}{\begin{eqnarray}}
\newcommand{\eeqa}{\end{eqnarray}}
\newcommand{\ba}{\begin{array}}
\newcommand{\ea}{\end{array}}

\begin{document}

\title{Zero Sound and First Sound in a Disk-Shaped Normal Fermi gas}
\author{Giovanni Mazzarella$^{1}$, Luca Salasnich$^{2}$
and Flavio Toigo$^{1}$}
\affiliation{$^{1}$Dipartimento di Fisica ``Galileo Galilei'' and CNISM,
Universit\`a di Padova, Via Marzolo 8, 35131 Padova, Italy \\
$^{2}$CNR-INFM and CNISM, Unit\`a di Padova, Via Marzolo 8,
35131 Padova, Italy}

\date{\today}

\begin{abstract}
We study the zero sound and the first sound in a
dilute and ultracold disk-shaped normal Fermi gas
with a strong harmonic confinement along the axial direction and uniform
in the two planar directions. Working at zero temperature
we calculate the chemical potential $\mu$ of the fermionic fluid
as a function of the uniform planar density $\rho$
and find that $\mu$ changes its slope
in correspondence to the filling of harmonic axial modes (shell effects).
Within the linear response theory, and under
the random phase approximation, we calculate
the velocity $c^{0}_s$ of the zero sound.
We find that also $c^0_s$ changes its slope
in correspondence of the filling of the harmonic axial modes
and that this effect depends on the Fermi-Fermi scattering length $a_F$.
In the collisional regime, we calculate the velocity $c_s$ of first sound
showing that $c_s$ displays jumps
at critical densities fixed by the scattering length $a_F$.
Finally, we discuss the experimental achievability of these zero sound
and first sound waves with ultracold alkali-metal atoms.
\end{abstract}


\maketitle

\section{Introduction}

Effects of quantum statistics are
observable with both bosonic \cite{book-pethick,book-stringari}
and fermionic \cite{demarco,truscott,modugno,greiner,joachim}
vapors of alkali-metal atoms at ultra-low temperatures.
In these quantum degenerate gases,
the role of dimensionality has been experimentally studied with
bosons \cite{gorlitz,schreck,kinoshita}, while for fermions, there are
only theoretical predictions. In particular, it has been suggested
that a reduced dimensionality strongly modifies density profiles
\cite{schneider,sala0,sala1,vignolo}, collective modes
\cite{bruun,minguzzi} and stability of mixtures \cite{das,sala2}.
Also sound velocities have been theoretically investigated in reduced
dimensions for both normal \cite{minguzzi,capuzzi1} and
superfluid Fermi gases \cite{ghosh,capuzzi2,capuzzi3}.

It has been shown that, in the weak coupling limit, the zero sound velocity
$c_s^0$ and the first sound velocity $c_s$ of a normal Fermi
gas are equal in the strictly one-dimensional (1D)
geometry \cite{minguzzi}, contrary to the 3D case where
they differ by a factor $\sqrt 3$ if the system is uniform
\cite{landau,pines,fetter},
or by a factor $\sqrt 5$ if the system is cigar-shaped due to a
isotropic harmonic confinement in two directions \cite{capuzzi1}.
Very recently we have studied the first sound velocity $c_s$
of a normal cigar-shaped Fermi gas in the 1D to 3D crossover, showing
that $c_s$ exhibits jumps in correspondence of the filling
of planar harmonic shells \cite{sala-reduce1}.

In this paper we investigate, at zero temperature,
the zero sound velocity $c_s^0$ and the first sound
velocity $c_s$ of a normal Fermi gas in the dimensional
crossover from 2D to 3D by
considering a harmonic confinement in the axial direction.
We show that by increasing the planar density $\rho$ of fermions
(or equivalently the chemical potential $\mu$) one
induces the dimensional crossover in the system. In general,
collective sound modes in a normal Fermi system are of two kinds.
When the collision time $\tau$ between quasiparticles is much greater
than the period
($\omega^{-1}$) of the mode itself, i.e. if $\omega \tau \gg 1$,
the system is in the so-called collisionless regime.  In presence of a
repulsive interaction, the
resulting collective mode is known as zero sound
\cite{landau,pines,fetter,negele,vignale,lipparini}.
In the present work we approach the problem calculating the
density-density response function in the random-phase approximation.
The poles of this
response function lead to the zero sound dispersion law and to its
velocity of propagation $c_s^0$ in the radial plane, where the gas is
not confined. We investigate the behavior of the zero
sound velocity $c_s^0$ as a
function of the chemical potential, or, equivalently, of the fermionic
density in the transverse radial plane.

When collisions between atoms
are characterized by a sufficiently small collision time
such that $\omega \tau \ll 1$,
the system may be assumed to be at thermal equilibrium. In this regime,
the Fermi fluid is accurately described by quantum hydrodynamics equations
where the fermionic nature appears in the bulk equation of state
\cite{landau,fetter,negele}. In this case, the collective mode is
called first sound (or ordinary sound, or collisional sound).
The first sound velocity $c_s$ has been studied previously
both in the 1D-3D and
2D-3D crossovers \cite{sala-reduce2,sala-reduce3}, and in a 3D cylindrical
geometry \cite{zaremba} for an ultracold Bose gas.
It is important to stress that the propagation of ordinary sound in
spin unpolarized Fermi gases may be
studied also from the experimental point of view. For instance, Joseph and
co-workers \cite{joseph} were able to excite first sound waves in an optically
trapped Fermi degenerate gas of spin-up and spin-down atoms by using
magnetically tunable interactions.
In this paper we predict that, as in the
1D-3D crossover \cite{sala-reduce1}, the first sound
velocity $c_s$ displays observable shell effects in
the 2D-3D crossover of a disk-shaped ultracold Fermi gas. By increasing the
planar density $\rho$, that is by inducing the crossover from two-dimensions to
three-dimensions, $c_s$ shows jumps in correspondence
of the filling of harmonic modes. Such jumps of the sound velocity,
even if less pronounced than the ones predicted in the
1D-3D crossover \cite{sala-reduce1}, can be experimentally observed.

In the last part of the paper we establish the experimental
conditions to be achieved for detecting zero sound and first sound.
In particular,
by considering $^{40}$K atoms, we estimate the critical
mode frequency $\omega_c$ which
discriminates between the collisional and the collisionless regimes.

\section{Confined interacting Fermi gas}

We consider an ultracold normal Fermi gas with two-spin components
confined by a harmonic trap in the axial direction and
investigate the regime where the temperature $T$ of the gas
is well below the Fermi temperature $T_F$ so that we may use a
zero-temperature
approach and treat separately
the collisionless and the
collisional regimes. The Hamiltonian of the dilute Fermi gas
trapped by the external harmonic potential
\beq
U({\bf r})=\displaystyle{\frac{1}{2} m\omega_z^2 z^2} \;,
\eeq
and free to move in the $(x,y)$ plane in a square box of size $L$,
is given by
\beqa
\label{hamiltonian}
&&\hat{H}=
\sum_{\sigma=\uparrow,\downarrow}\int d {\bf r}
\hat{\Psi}^{\dagger}_{\sigma}({\bf r})\bigg(-\frac{\hbar^2}{2m}
\nabla^2+U({\bf r})\bigg)\hat{\Psi}_{\sigma}({\bf r})\nonumber\\
&+&\frac{g_{F}}{2}\sum_{\sigma,\sigma'=\uparrow,\downarrow}\int d
{\bf r} \hat{\Psi}^{\dagger}_{\sigma}({\bf
r})\hat{\Psi}^{\dagger}_{\sigma^{'}}({\bf
r})\hat{\Psi}_{\sigma^{'}}({\bf r})\hat{\Psi}_{\sigma}({\bf r})
(1-\delta_{\sigma,\sigma'}) \;,\nonumber
\eeqa
where $\sigma=\{\uparrow,\downarrow \}$ denotes the spin variable,
and the inter-atomic coupling is
\beq
g_{F}=4 \pi \hbar^{2} a_{F}/m \; ,
\eeq
with $a_F>0$ being the $s$-wave scattering length.
${\hat \Psi}_{\sigma}({\bf r})$ and
${\hat \Psi}_{\sigma}^{\dagger}({\bf r})$
are the usual operators
annihilating or creating a fermion of spin $\sigma$ at the
position ${\bf r}$, therefore obeying anticommutation relations
\cite{landau,pines,fetter,negele}.
The field operator can be
written as \beq {\hat \Psi}_{\sigma}({\bf r})= \sum_{\alpha}
\phi_{\alpha}({\bf r}) \, {\hat c}_{\sigma, {\alpha}} \eeq in
terms of the single-particle operators ${\hat c}_{\sigma, \alpha}$
(${\hat c}^{\dagger}_{\sigma, \alpha}$) destroying (creating) a
fermion of spin $\sigma$ with the single-particle wave function
$\phi_{\alpha}({\bf r})$.
Because of the symmetry of the problem,
$\phi_{\alpha}({\bf r})$ can be factorized into the product of
eigenfunctions of the harmonic oscillator in the $z$ direction
and plane waves describing free motion in the $(x,y)$ plane.
The field operator, then, may be written as
\beqa
\label{fieldoperator} \hat{\Psi}_{\sigma}({\bf r})=\frac{1}{L^2}
\sum_{j,{\bf K}} \psi_{j}(z) \exp(i {\bf K} \cdot {\bf R})
\hat{c}_{\sigma,j,{\bf K}} \;
\eeqa
with ${\bf R} \equiv (x,y)$,
and where $\psi_{m}(z)$ is the real function
\beq
\label{eigenfunctions}
\psi_{j}(z)=\sqrt{\frac{1}{2^{j} j! \pi^{1/2} a_z}}
H_{j}(\frac{z}{a_z}) \exp(-z^2/2a_z^2) \; .
\eeq
$\displaystyle{H_{j}(\frac{z}{a_z})}$ is the $j$-th Hermite
polynomial of argument $z/a_z$, $a_z=\sqrt{\hbar/(m \omega_z)}$
being the characteristic length of the axial harmonic potential.

The two dimensional wave vector ${\bf K}\equiv (K_x,K_y)$
embeds the translational
invariance in the radial plane. By imposing periodic boundary
conditions in the box of length $L$ along $x$ and $y$ directions,
the components $K_x$ and $K_y$ of ${\bf K}$ are quantized according
to: $\displaystyle{K_x=\frac{2 \pi}{L}{i_x}}$ and
$\displaystyle{K_y=\frac{2 \pi}{L}{i_y}}$, where $i_x$ and $i_y$
are integer quantum numbers. The index
$\alpha$ reads $(j, K_x, K_y )$, or in a more compact form
$(j, {\bf K} )$.

By inserting Eq. (\ref{fieldoperator}) into (\ref{hamiltonian})
the Hamiltonian takes the form
\beq
\label{chamiltonian}
\hat{H} = \hat{H}_0 + \hat{H}_{int} \;,
\eeq
where
\beq
\hat{H}_0= \sum_{\sigma,j,{\bf K}}\epsilon^{0}_{j,{\bf K}}
\hat{c}^{\dagger}_{\sigma,j,\bf K} \hat{c}_{\sigma,j,\bf K} \;,
\eeq
and
\beqa
\label{chamiltonian1}
&&\hat{H}_{int}=\frac{g_{F}}{2 a_z L^2}
\sum_{\sigma,\sigma'}\sum_{j_{1},j_{2},j_{3},j_{4}}
\sum_{\bf K_1,\bf K_2,\bf Q}
W_{j_{1},j_{2},j_{3},j_{4}}
\nonumber
\\
&& \hat{c}^{\dagger}_{\sigma,j_1,\bf K_1+\bf Q}
\hat{c}^{\dagger}_{\sigma',j_2,{\bf K_2}-\bf
Q}\hat{c}_{\sigma',j_3,{\bf K_2}}\hat{c}_{\sigma,j_4,{\bf K_1}}
(1-\delta_{\sigma,\sigma'}) \;, \nonumber\\ \eeqa with \beq
\label{singleparticleenergy} \epsilon^{0}_{j,\bf K}=\frac{\hbar^2
K^{2}}{2m} +(j+\frac{1}{2}) \hbar \omega_{z} \;,  \eeq
and
\beqa
&&W_{j_1,j_2,j_3,j_4} =
\bigg(\frac{1}{\pi}\prod_{j_i,i=1}^4\sqrt{\frac{1}{2^{j_i}
j_{i}!}}\bigg)\nonumber\\
&&\int_{-\infty}^{+\infty} d \xi e^{-2 \xi^2}
H^{*}_{j_1}(\xi)H^{*}_{j_2}(\xi)H_{j_3}(\xi)H_{j_4}(\xi)
\label{wal}.
\eeqa

Eq. (\ref{singleparticleenergy}) gives the single-particle
energies of noninteracting fermions. To obtain the single-particle
energies of interacting fermions we calculate
the two-spin Green function
\beq
\label{green}G^{\sigma,\sigma'}_{j_{1},j_{2},{\bf K}}(t-t')
=-\frac{i}{\hbar}\langle T \big(\hat{c}_{\sigma,j_1,\bf K}
(t)\hat{c}^{\dagger}_{\sigma',j_2,\bf K}(t')\big)\rangle \;
\eeq
where $T$ denotes the time ordering operator. Here and in the following,
we will represent the operators in the interaction picture (with
$\hat{H}_0$ the unperturbed Hamiltonian), and will evaluate the
averages with respect to the ground state of the Hamiltonian
(\ref{chamiltonian}). In the calculation of the Green function
we use a mean field approach and replace the four operators terms involved
in $\hat{H}_{int}$
with
$\langle\hat{c}^{\dagger}_{\sigma,j_1,\bf K_1+\bf
Q}\hat{c}_{\sigma,j_4,{\bf K_1}}\rangle
\hat{c}^{\dagger}_{\sigma',j_2,\bf K_2-\bf
Q}\hat{c}_{\sigma',j_3,{\bf
K_2}}+\hat{c}^{\dagger}_{\sigma,j_1,\bf K_1+\bf
Q}\hat{c}_{\sigma,j_4,{\bf K_1}} \langle
\hat{c}^{\dagger}_{\sigma',j_2,{\bf K_2}-\bf
Q}\hat{c}_{\sigma',j_3,{\bf K_2}} \rangle $.
The poles of the Green function with $\sigma=\sigma'$ and $j_{1}=j_{2}=j$
give directly
the single-particle energies $\epsilon_{j,{\bf K}}$ of the interacting
fermions
\beqa
\label{spwi}
\epsilon_{j,{\bf K}}=\frac{\hbar^2 K^{2}}{2m} +(j+\frac{1}{2})
\hbar \omega_{z}+\frac{g_F}{2 a_z  L^2
}\;\sum_{l=0}^{l_{max}} u_{j,l} {n}_l
\nonumber\\
\eeqa
where
\beq
{n}_l=
\sum_{{\bf K}, \sigma}
\langle\hat{c}^{\dagger}_{\sigma,l,{\bf K}}
\hat{c}_{\sigma,l,{\bf K}}\rangle \;
\eeq
is the total number of fermions in the harmonic shell $l$ and $u_{j,l}
=u_{l,j}=W_{l,l,j,j}$ is obtained from Eq. (\ref{wal}).

In Tab. 1 we report the numerical values of the coefficients $u_{l,j}$ for
$l$ and $j$ up to $4$.

The maximum positive integer $l_{max}$ involved in the summation at the
right hand side of Eq. (\ref{spwi}) clearly depends on the single-particle
state $(j,{\bf K})$ and on the interaction strength $g_F$.

If spin-up and spin-down states are equally
populated we may write the total number of fermions in the $l$-th harmonic
state as
\beq \label{total} {n}_l = 2\, \sum_{{\bf K}} \,
\Theta( \bar \mu - \epsilon_{l,{\bf K}} ) \; .
\eeq
with $\bar \mu$ the chemical potential, i.e. the Fermi energy.

\begin{figure}\centering
\includegraphics[width=8cm,clip]{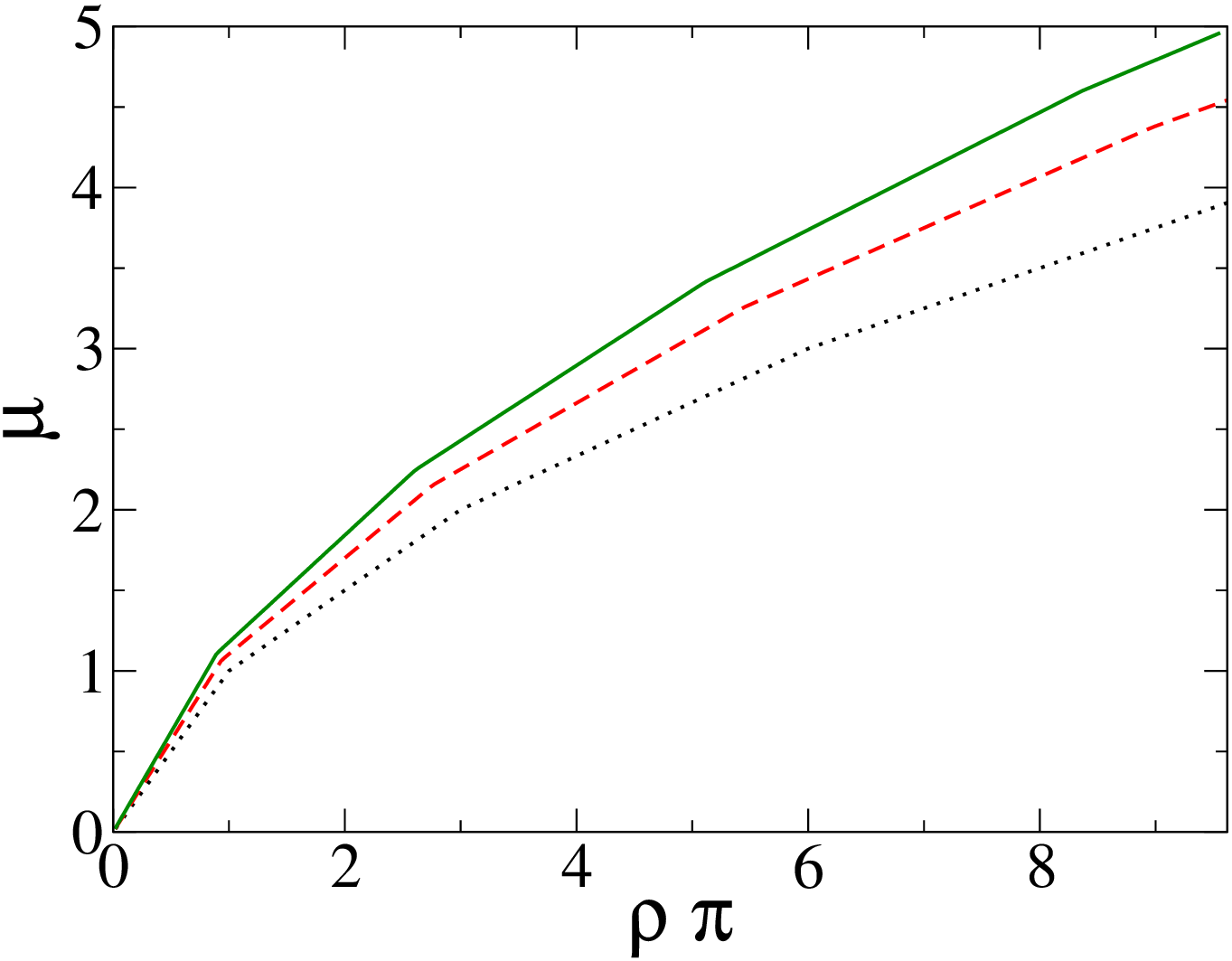}
\caption{(Color online).
Chemical potential $\mu$ vs. planar density $\rho$.
Three values of the scaled interaction strength:
$g=10^{-5}$ (dotted curve), $g=0.35$ (dashed curve),
$g=0.6$ (solid curve).
The chemical potential is in units of $\hbar \omega_z$, density in
units of $1/a_z^2$, and lengths in units of $a_z$.}
\label{Fig1}
\end{figure}

From here we may calculate the 2D fermion density in the $(x,y)$
plane as a function of the chemical potential.
Under the condition $L \gg a_z$
on the planar box size, the summation over $\bf K$ may be
repaced by an integral to get
the 2D density associated to the $j$th axial mode, which
is uniform in the $(x,y)$ plane and is given by
\beq{\label{nj}}
\rho_j = {{n}_j\over
L^2} = 2 \frac{1}{(2 \pi)^2}  \int d^2 {\bf K}   \, \Theta( \bar \mu -
\epsilon_{j,{\bf K}} ) \; .
\eeq

\begin{center}
\begin{tabular}{|c|c|}
\hline
~~~$(l,j)$~~~ & ~~~$u_{l,j} \ \sqrt{2\pi}$~~~ \\
\hline
 $(0,0)$ & $1$    \\
 $(0,1)$ & ${1/ 2}$  \\
 $(0,2)$ & ${3 / 8}$ \\
 $(0,3)$ & ${5 /16}$ \\
 $(0,4)$ & ${35 / 128}$ \\
 $(1,1)$ & ${3 / 4}$ \\
 $(1,2)$ & ${7 / 16}$ \\
 $(1,3)$ & ${11 / 32}$ \\
 $(1,4)$ & ${75 / 256}$ \\
 $(2,2)$ & ${41 / 64}$ \\
 $(2,3)$ & ${51 / 128}$ \\
 $(2,4)$ & ${329 / 1024}$ \\
 $(3,3)$ & ${147 / 256}$  \\
 $(3,4)$ & ${759 / 2048}$ \\
 $(4,4)$ & ${8649 / 16384}$ \\
\hline
\end{tabular}
\end{center}
\noindent Table 1.
{\small Coefficients $u_{l,j}$. Note that $u_{l,j}=u_{j,l}$.}

\vskip 0.5cm

If only only the lowest ($j=0$)
single-particle mode along the $z$ axis is occupied, then the system may be
considered as strictly 2D.
After inserting Eq. (\ref{spwi}) into Eq. (\ref{nj})
an easy integration gives
\begin{eqnarray}
\rho_j=\frac{ m}{\pi \hbar^2} (\mu-\hbar \omega_z j-
\frac{g_F}{2a_z} \sum_{l=0}^{l_{max}} u_{j,l}\rho_{l}) \; ,
\label{pippa}
\end{eqnarray}
where
\beq
\mu=\bar \mu -\frac{\hbar \omega_z}{2} \;
\eeq
is the chemical potential measured with respect to the ground state energy
of the axial harmonic potential.
The expression (\ref{pippa}) may be rewritten as
\beq
\label{nvsmu}
\sum_{l=0}^{l_{max}}\big(\delta_{j,l}+g u_{j,l}\big)\rho_{l}
=\frac{m}{\pi \hbar^2}(\mu-\hbar \omega_z j) \; ,
\eeq
where
\beq
g= \label{gbarra}\frac{g_F}{2 \pi \hbar \omega_z a_z^3}
= {2 a_F\over a_z} \:
\eeq
is the scaled inter-atomic strength.

The total planar density is then given by
\beq
\rho = \sum_{l=0}^{l_{max}} \rho_{l} \;,
\eeq
where the only nonzero densities $\rho_0 , \rho_1 ,...,\rho_{l_{max}-1},
\rho_{l_{max}}$ are obtained
by solving the ${l_{max}}+1$ Eqs. (\ref{nvsmu}) at fixed $\mu$ and $g$.
The value of $l_{max}$   is  determined by requiring that all the $l_{max}+1$
solutions are positive.
Note that when only $\rho_0$ is non zero, we have a strictly 2D system.
In Fig.1 we plot the chemical potential $\mu$ as a function
of the planar density $\rho$ for three values of the
scaled interaction strength $g$. The figure shows the
changes in the slope of these curves which occur when
the planar density is sufficiently large so that
a new excited level of the axial harmonic well (a new shell)
begins to be populated.

\section{Zero sound}

In the present section we analyze the collective
mode of the fermionic gas in the collisionless regime, i.e  zero sound.
To this end we calculate the density-density response function:
\beqa
\label{responsef} &&\chi^{\sigma,\sigma}({\bf Q},t)=
\nonumber\\
&& -\frac{i}{\hbar} \Theta(t)<[\hat{n}_{\sigma}({\bf
Q},t), \hat{n}^{\dagger}_{\sigma}({\bf Q,0})]>
\label{respf}
\eeqa
since its poles provide the dispersion relation and the velocity
of propagation of this mode.

In Eq. (\ref {respf}) we have introduced the 2D
density number operator $\hat{n}_{\sigma}(\bf Q)$ according to
\beq
\hat{n}_{\sigma}({\bf Q})
\equiv
\sum_{j,{\bf K}} \hat{n}_{\sigma,j,{\bf K}} ({\bf Q})
= \sum_{j,{\bf K}} \hat{c}^{\dagger}_{\sigma,j,{\bf K}}
\hat{c}_{\sigma,j,{\bf K}+{\bf Q}} \; .
\eeq
At any time, we may define
 \beq \hat{n}_{\sigma,j}({\bf Q},t)=
\sum_{{\bf K}}\hat{n}_{\sigma,j,{\bf K}}({\bf Q},t) \; .
\eeq

We proceed by calculating the equations of motion for spin diagonal
($\sigma = \sigma'$) and non diagonal ($\sigma \neq \sigma'$)
density-density response functions
\beqa
\label{motionequations} && i \hbar \frac{d}
{d t}\chi^{\sigma,\sigma}_{j,j,{\bf K}}({\bf Q},t)\nonumber\\
&=&\delta(t)<[ \hat{n}_{\sigma,j,{\bf K}}({\bf Q},t),
\hat{n}^{\dagger}_{\sigma,j}({\bf Q},0)]> \nonumber\\
&-&\frac{i}{\hbar} \Theta(t)<[
[\hat{n}_{\sigma,j,{\bf K}}({\bf Q},t), \hat{H}(t)],
\hat{n}^{\dagger}_{\sigma,j}({\bf Q},0)]>,\nonumber\\
&& i \hbar \frac{d}{d t}\chi^{\sigma,\sigma'}_{j,j,{\bf K}}({\bf Q},t)
\nonumber\\
&=&\frac{i}{\hbar} \Theta(t)<[ [\hat{n}_{\sigma,j,{\bf K}}({\bf
Q},t), \hat{H}(t)], \hat{n}^{\dagger}_{\sigma',j}({\bf Q},0)]>.
\nonumber
\eeqa

After a lengthy but straightforward calculation and
using  RPA \cite{pines,vignale} to decouple the higher order response
function coming from the commutator on the r.h.s. of the equation of motion,
the time Fourier transforms $\chi_{l,l}({\bf Q},\omega)$ of  the
spin diagonal and non-diagonal density-density response
functions read, respectively
\beqa
\label{chis} \chi^{\sigma,\sigma}_{j,j}(\bf Q,\omega) &=&
\chi^{(0)}_{j,j}({\bf
Q},\omega)\nonumber\\
&+&\frac{g_{F}}{a_z L^2} \chi^{(0)}_{j,j}({\bf Q},\omega)
\sum_{l=0}^{l_{max}} u_{j,l} \chi^{\sigma',\sigma}_{l,l}({\bf Q},\omega)
\nonumber\\
\chi^{\sigma',\sigma}_{j,j}(\bf Q,\omega) &=& \frac{g_{F}}{a_z L^2}
\chi^{(0)}_{j,j}({\bf Q},\omega)
\sum_{l=0}^{l_{max}} u_{j,l} \chi^{\sigma,\sigma}_{l,l}({\bf Q},\omega)\;.
\nonumber
\eeqa
The function $\chi^{(0)}_{j,j}({\bf Q},\omega) \equiv
\chi^{(0),\sigma,\sigma}_{j,j}({\bf Q},\omega)$, appearing in both
equations, is given by
\beq
\label{chizero} \chi^{(0)}_{j,j}({\bf Q},\omega)=
\sum_{\bf K} \frac{n_{\sigma,j,\bf K}-n_{\sigma,j,\bf K+\bf Q}}
{\hbar \omega -(\epsilon^{0}_{j,\bf K+\bf Q}-\epsilon^{0}_{j,\bf K})
+i \hbar \eta},
\eeq
where the positive imaginary term in the denominator ensures causality
and the limit $\eta \rightarrow 0^{+}$ will eventually be taken.
The totally symmetric density-density response function defined as
\beqa
&&\chi({\bf Q},\omega)=
\frac{1}{4}\nonumber\\
&&\big[\chi^{\uparrow,\uparrow}({\bf
Q},\omega)+\chi^{\downarrow,\uparrow}({\bf
Q},\omega)+\chi^{\uparrow,\downarrow}({\bf Q},\omega)+
\chi^{\downarrow,\downarrow}({\bf Q},\omega)\big]\nonumber\\
\eeqa finally reads
\beqa \chi_{j,j}(\bf Q,\omega) &=& \frac
{\chi^{(0)}_{j,j}(\bf Q,\omega)}{2}
\nonumber\\
&+&\frac{g_{F}}{a_z L^2} \chi^{(0)}_{j,j}({\bf
Q},\omega)\sum_{l=0}^{l_{max}} u_{j,l} \chi_{l,l}({\bf Q},\omega)
\label{equazionefinale}\;.\nonumber\\
\eeqa This equation is one of the main results of the paper. It
may be recast in the form of a system of equations for the
response functions $\chi_{l,l}({\bf Q},\omega)$ pertaining to the
various axial modes $l$: \beq \sum_{l=0}^{l_{max}}
\bigg(\delta_{j,l}-\frac{g_{F}}{a_z L^2} u_{j,l}
\chi^{(0)}_{j,j}({\bf Q},\omega)\bigg) \chi_{l,l}({\bf Q},\omega)=
\frac {\chi^{(0)}_{j,j}(\bf Q,\omega)}{2} \; \label{recast} \eeq
in terms of which the full response function is given by \beq
\chi({\bf Q},\omega)=\sum_{l=0}^{l_{max}} \chi_{l,l}({\bf
Q},\omega) \; \label{chiff} \eeq

The poles of this response function coincide with the zeros of the
determinant of the matrix of coefficients of $\chi_{l,l}({\bf
Q},\omega)$ in the system of equations (\ref{recast}) in the limit
$\eta \to 0^{+}$. Since zero sound  is appreciably damped for
large ${ Q}$, \cite{pines}, we look for zeros of the determinant
in the limit  ${ Q} \to 0$. In this regime, one may evaluate
$\chi^{(0)}_{l,l}({\bf Q},\omega)$ of Eq.(\ref{chizero}) under the
condition \beq s= \frac {\omega}{ Q} > \sqrt {2 \frac{\pi
\hbar^2}{m^2} \rho_l} \label{sl} \eeq to find \beq
\chi^{(0)}_{l,l}({\bf Q},\omega ) = \frac{m L^2}{2 \pi \hbar^2}
\Big (\frac{s}{\sqrt {s^2-2 \frac{\pi \hbar^2}{m^2} \rho_l}} \Big
) \label{chi0} \; ,\eeq where use of Eq. (\ref{pippa}) has been
made. Note that
\beq v_{F,l} = \sqrt {2 \frac{\pi \hbar^2}{m^2}
\rho_l} \label{vf0} \eeq
is the planar Fermi velocity in the
$l$-th axial shell ($l=0,1,...,l_{max}$). $v_{F,l}$ corresponds to
the Fermi velocity of a strictly 2D uniform Fermi gas with density
$\rho_l$, which at fixed $\mu$ is a function of $g$.
Using Eq. (\ref{chiff}), we may calculate the response of the
system for any value of the chemical potential, i.e. for any
number of  excited levels of the axial harmonic trap. We start by
considering the case when only the lowest state of the harmonic
axial well is populated, i.e. $l_{max}=0$. Then, from  Eq.
(\ref{recast}), one gets the density response function
$\chi_{0,0}({\bf Q},\omega)$ as \beq \label{chigs} \chi_{0,0}({\bf
Q},\omega)= \frac{\chi^{(0)}_{0,0}({\bf Q},\omega)/2} {1- {(g_F/
a_z L^2)}\, u_{0,0}\chi^{(0)}_{0,0}({\bf Q},\omega)}\:, \eeq

By using Eq. (\ref{chi0}) with $l=0$, we find easily that the poles of Eq.
(\ref{chigs}) satisfy $\omega = c_s^{0} Q$ where the velocity of zero sound
is given by
\beq
\label{zerosoundvelocity}
c_s^{0} = v_{F,0}  \ f(g)
\eeq
with
\begin{eqnarray}
\label{fbarg}
f(g)= {1 + g \ u_{0,0} \over
\sqrt{1 + 2\ g\ u_{0,0} } } \; ,
\end{eqnarray}
the term which takes into account the effects of interaction. We
remind that the chemical potential $\mu$ is related to the 2D
density $\rho$ by Eq. (\ref{nvsmu}), i.e. \beq \label{muvsnzero}
\mu = {{\pi \hbar^2}\over m}  \ (1+g \ u_{0,0} )\ \rho \; . \eeq
The expression (\ref{zerosoundvelocity}) for the zero sound
velocity coincides with that of a strictly 2D Fermi gas
\cite{lipparini} of density $\rho_0$, with the effect of the
axial harmonic trap embodied in the definition of the
dimensionless parameter $g \ u_{0,0}$ as the effective 2D
coupling. We recognize in this effective 2D coupling
the familiar Landau parameter $F_{0}$ \cite{lipparini} for a strictly
interacting 2D Fermi gas. Let us denote $g \ u_{0,0}$ by $F_{0}$. By
using this definition and the Eq. (\ref{muvsnzero}), we may
express the zero sound velocity (\ref{zerosoundvelocity}) in terms
of $F_{0}$ 
\beq \label{zerosoundvelocitybis} c_s^{0} =
\sqrt{\frac{2}{m}(\mu-\pi \hbar \omega_z a_z^2 \rho F_{0}
)}
 {1 + F_{0} \over
\sqrt{1 + 2\ F_{0} } }
.\eeq

\begin{figure}\centering
\includegraphics[width=8cm,clip]{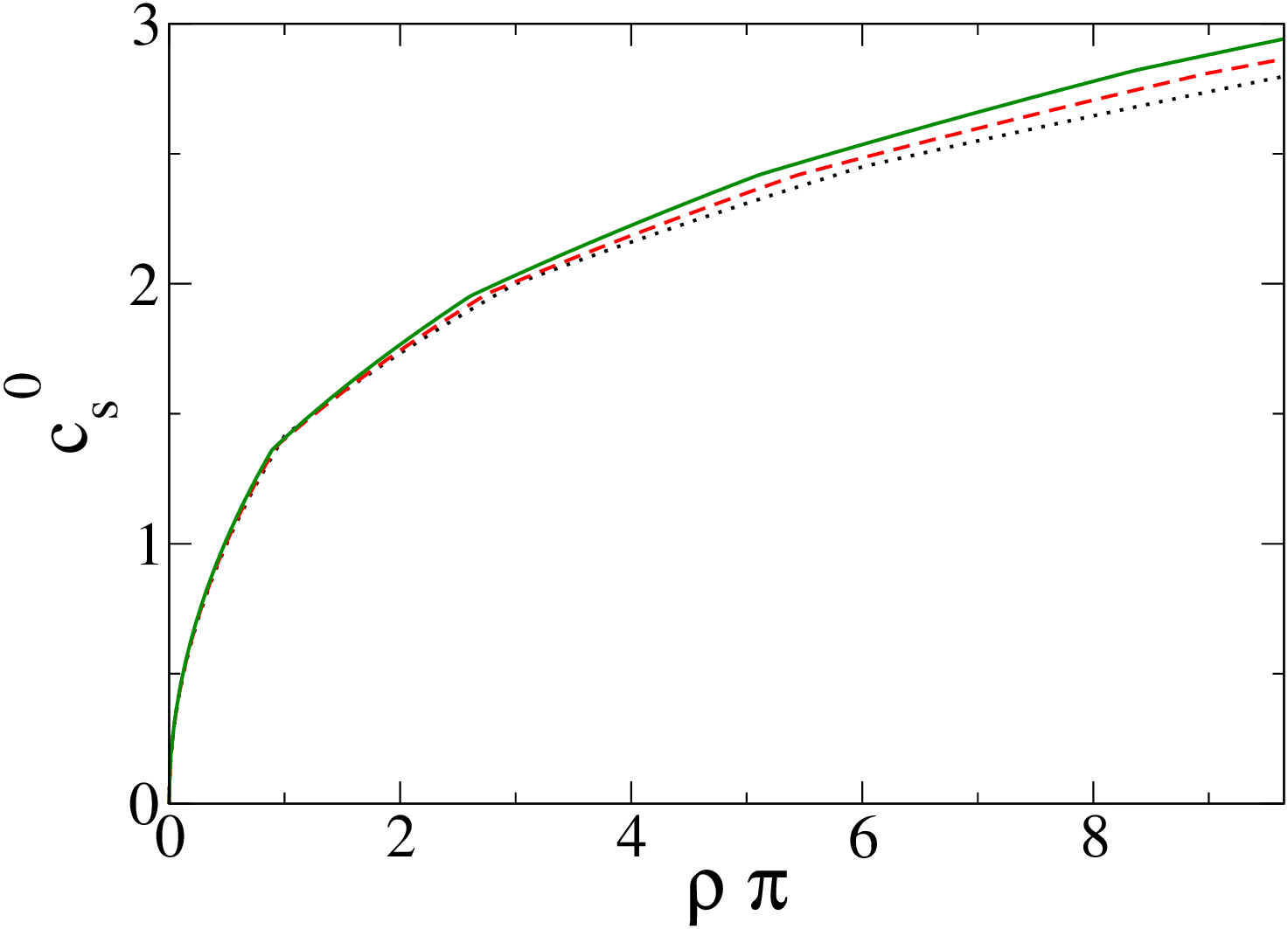}
\caption{(Color online).
Zero sound velocity $c_s^{0}$ vs. planar density $\rho$.
Three values of the scaled interaction strength:
$g=10^{-5}$ (dotted curve), $g=0.35$ (dashed curve),
$g=0.6$ (solid curve).
The zero sound velocity is in units of
$a_z \omega_z$, density in units of $1/a_z^2$, and lengths in units of $a_z$.}
\label{Fig2}
\end{figure}

Now we consider a density $\rho$
large enough so that some excited axial harmonic levels are populated.
In this case, the velocity of zero sound is given by the zero of
the determinant
of the matrix in the l.h.s. of Eq. (\ref{recast}) with $\omega > Q$.
The calculation is again straightforward and the results are reported
in Fig. 2, where we display the
zero sound velocity $c_s^{0}$ as a function of the planar density $\rho$.
As it happens in the strictly 2D case,
for a given value of the planar density $\rho$, $c_s^{0}$ increases
by increasing the repulsive two body interaction $g$.

\begin{figure}\centering
\includegraphics[width=8cm,clip]{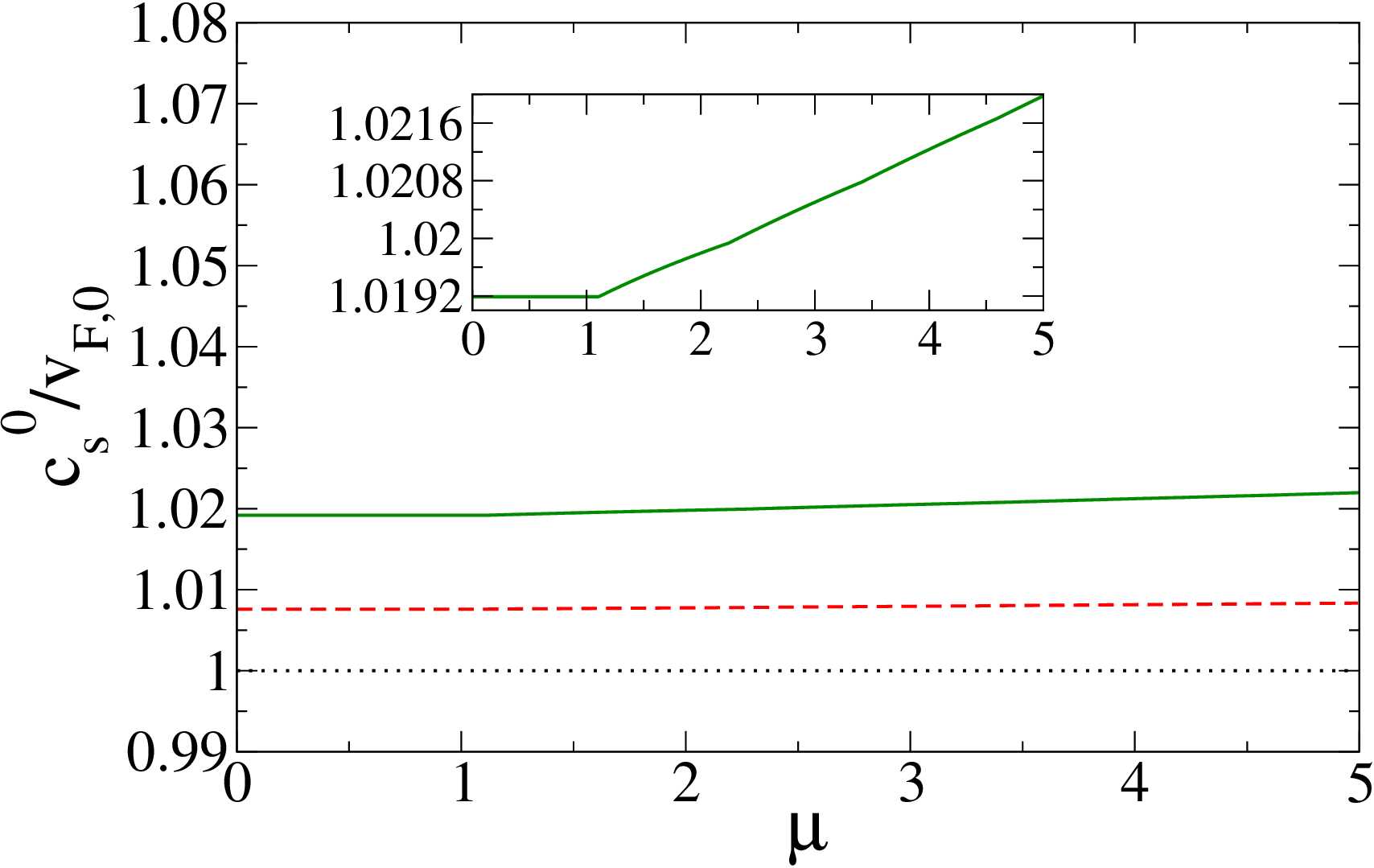}
\caption{Ratio $c_s^{0}/v_{F,0}$ vs. chemical potential $\mu$.
Three values of the scaled interaction strength:
$g=0$ (dotted curve), $g=0.35$ (dashed curve),
$g=0.6$ (solid curve).
Note that the dotted curve is superimposed
to the solid one. The velocities $c_s^0$ and $v_{F,0}$ are in units of $a_z
\omega_z$, lengths in units of $a_z$, and the chemical potential in units of $\hbar \omega_z$.}
\label{Fig3}
\end{figure}

In Fig. 3 we plot the quantity $c_s^{0}/v_{F,0}$
as a function of the chemical potential $\mu$.
When $g=0$ one finds that $c_s^{0}$=$v_{F,0}$. If $g>0$ but $\mu$ is
sufficiently small so that  only the lowest axial mode is
populated, then the quantity $c_s^{0}/v_{F,0}$ does not change when the
chemical potential is varied (dashed line).
This behavior corresponds to the one
described by Eq. (\ref{zerosoundvelocity}). When the chemical
potential is greater than a critical value $\mu_c$,
excited axial modes begin to be occupied
by the fermions and $c_s^{0}/v_{F,0}$ grows with $\mu$ (see the
inset of Fig. 3).

\section{First sound}

In this section we analyze the collisional case of a normal
paramagnetic Fermi gas.
To analyze a sound wave that travels in a planar direction it is
important to determine the collision time $\tau$ of the gas.
According to \cite{bruun} and \cite{gupta}, if the local Fermi
surface is not strongly deformed then $\tau = \tau_0 (T/T_F)^2$,
where $\tau_0=1/(\rho^{(3D)} \sigma v_F)$ with $\sigma$ the scattering
cross-section and $\rho^{(3D)}$ the 3D density. Instead, if the local
Fermi surface is strongly deformed, as in our case, then the
collision time is simply given by $\tau=\tau_0$ \cite{gupta}. It
is easy to find that in our 2D geometry $\tau_0 \sim 1/(\omega_z \rho
a_F^2)$. Setting $\omega$ the frequency of oscillation of the
sound wave, the wave propagates in the collisionless regime if
$\omega \tau \gg 1$ and in the collisional regime if $\omega \tau
\ll 1$.

In the collisional regime by adopting the
equations of hydrodynamics \cite{landau,fetter,negele} the planar
first sound velocity $c_s$ of the Fermi gas is given by the
zero-temperature formula
\beq
c_s = \sqrt{ { \rho \over m} {\partial
\mu\over \partial \rho}} \; . \label{sound-c2}
\eeq
By using this formula and Eq. (\ref{nvsmu}) we determine the
behavior of $c_s$ versus $\rho$ and of $c_s/v_{F,0}$ as a function of
$\mu$, as shown in Fig. 4 and in Fig. 5, respectively. These
figures, obtained both for $g=0$ and for $g> 0 $ display
shell effects, namely jumps of the first sound velocity $c_s$ when
the atomic fermions occupy a new axial harmonic mode.

If the interaction is negligible in the equation of state,
the values $\rho_{c}$ of the density at which such jumps occur
may be analytically obtained from Eq. (\ref{nvsmu}).
In fact, we obtain the behavior of the
planar density as a function of the chemical potential by setting
$g=0$ in the Eq. (\ref{nvsmu}) and performing the sum over the
quantum number $j$. Then, we get
\beq
\label{n2d-mu}
\rho =
\sum_{j=0}^{I[{\mu\over \hbar\omega_z}]} {1\over \pi a_z^2} \,
\left({\mu\over \hbar\omega_z} - j \right) = {1\over \pi a_z^2} \,
S({\mu\over \hbar\omega_z}) \;,
\eeq
where \beq S(x) = (1+I[x])
(x - {1\over 2} I[x]) \; \label{s2-exact}
\eeq
with $I[x]$ the integer part of $x$. Eq. (\ref{n2d-mu})
with Eq. (\ref{s2-exact})
represents a simple analytical formula which gives the uniform
planar density $\rho$ of the Fermi gas as a function of the chemical
potential $\mu$. The values $\rho_c$ signing the jumps can be
obtained from Eq. (\ref{n2d-mu}) with with Eq. (\ref{s2-exact}).
The jumps of the first sound velocity are given by
\beq
\rho_{c}=\displaystyle{\frac{k(k+1)}{2 \pi a_{z}^2}} \quad
\mbox{with} \quad k=1,2,3,... \; .
\eeq
When the critical value of
the density is approached from the left, from Eq. (\ref{sound-c2})
we find that the sound velocity is
\beq
c_s^{-}=\displaystyle{\omega_z a_z\sqrt{\frac{k+1}{2}}} \;,
\eeq
while when $n_{c}$ is approached form the right, we find
\beq
c_s^{+}=\displaystyle{\omega_z a_z\sqrt{\frac{k}{2}}} \; .
\eeq
The height of the first jump of $c_s$ is then given
by \beq \Delta c_s = \omega_z a_z\frac{\sqrt{k+1} -\sqrt{k}}
{\sqrt{2}} \; .
\eeq

When interactions are not negligible in the equation of state, one
must use Eq. (\ref{nvsmu}) to get the values of $\rho$ for which
the slope of $\mu$ exhibits discontinuities, and calculate the
corresponding discontinuities in $c_s$. For example, at the first
discontinuity, which appears when $l_{max}$ goes from $0$ to $1$
the value of the density is
\beq \rho_c=\frac{1}{\pi a_z^2}
\frac{1}{1 +g u_{0,0}}=\frac{1}{\pi a_z^2} \frac{1}{1 +F_{0}
} \; . \label{roc} \eeq
The corresponding discontinuity
exhibited by the first sound velocity is \beq \Delta c_s=
c_s^{-}-c_s^{+}\;, \eeq where \beq c_s^{-}=\omega_z
a_z\sqrt{\frac{1}{(1+ g u_{0,0})[1+ g (u_{0,0}-u_{1,1})]}} \;,
\eeq and

\beqa c_s^{+} &=& \omega_z a_z \nonumber\\
&&\sqrt{\frac{1+g [u_{0,0}+u_{1,1}+g (u_{0,0}
u_{1,1}-{u^2}_{0,1})]} {[1+ g (u_{0,0}-u_{1,1})]  [2+g (u_{0,0}-2
u_{0,1}+u_{1,1})]}}\;, \nonumber\\
\eeqa where the $u$'s are given in Tab. 1. We observe that these
shell effects are similar to the ones occurring in the 1D-3D
crossover \cite{sala-reduce1}, where the jumps are sharper because
the sound velocity goes to zero in correspondence of the filling
of planar harmonic modes, showing the crucial role played by the
dimensionality. In the next section we shall discuss the
experimental conditions necessary to observe shell effects in the
velocity of first sound.

\begin{figure}
\centering
\includegraphics[width=8cm,clip]{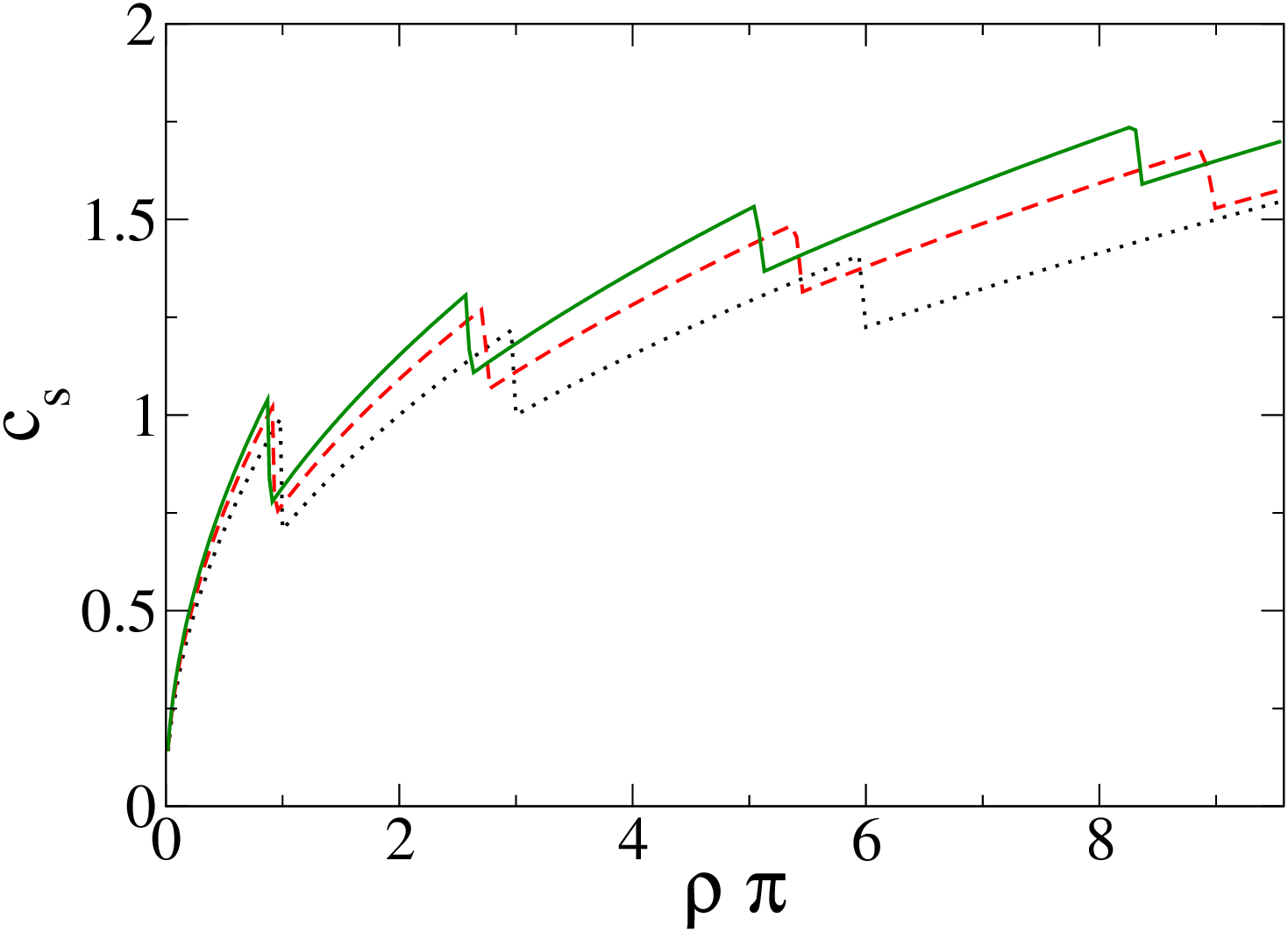}
\caption{(Color online). First sound velocity $c_s$ as a function
of the transverse density $\rho$.
Three values of the scaled interaction strength:
$g=0$ (dotted curve), $g=0.35$ (dashed curve),
$g=0.6$ (solid curve). The first sound velocity is in units of
$a_z \omega_z$, density in units of $1/a_z^2$, and lengths in units of $a_z$.}
\label{Fig4}
\end{figure}

Now we comment on the 2D-3D crossover in our fermionic gas,
starting from the strictly 2D case, corresponding to $l_{max}=0$
in Eq. (\ref{pippa}). In this case the chemical potential $\mu$ is
given by Eq. (\ref{muvsnzero}) and therefore from Eqs. (\ref{vf0})
and (\ref{sound-c2}) one finds 
\beq c_s={v_{F,0} \over
\sqrt{2}}\sqrt{(1+ g u_{0,0})}\; ,\eeq or, in terms of the Landau
parameter $F_{0}$ \beq \label{fs}
c_s=\frac{\sqrt{\frac{2}{m}(\mu-\pi \hbar \omega_z a_z^2 \rho
F_{0} )}} {\sqrt{2}}\sqrt{1+ F_{0}}\; \eeq
recovering the usual result for a 2D Fermi system. Such a behavior
is illustrated by the first plateau in Fig. 5. Incidentally, note
that when $F_{0} \simeq 0$ the zero sound velocity
(\ref{zerosoundvelocitybis}) is larger than the first sound
velocity (\ref{fs}) by a factor of $\sqrt{2}$. In our case, the quantity $F_{0}$ is
smaller than one. Then the two velocities are different between each other
also when $F_{0}$ remains finite.

We observe that the first sound velocity depends explictly on the
chemical potential. In fact, remembering the relation between
$\mu$ and $\rho$ given by Eq. (\ref{nvsmu}), and using Eqs.
(\ref{n2d-mu}) and (\ref{s2-exact}), one may show that the
velocity of first sound takes the form: \beq \label{csvsmu} c_s=
\sqrt{2\mu \over m}
 \sqrt{ {1 \over 2} \Big[ 1-{1 \over 2}
\frac{I[\mu / \hbar \omega_z]}{ \mu / \hbar \omega_z}+\frac{t(g)}{\mu /
\hbar \omega_z} \Big] } \; ,\eeq
where $t(g)$ is a complicated but analytical function of
$g$ and of the matrix elements $u$'s, which vanishes if $g =0$.
We stress that the chemical potential
$\mu$ depends on the planar density $\rho$ and on the
interaction strength $g$.
Fig. 5 shows how $c_s/v_{F,0}$ explictly depends on $\mu$  when
axial harmonic states other than the ground states are occupied.
From Eq. (\ref{csvsmu}) one finds that for any $g$ at very
large $\rho$ (i.e. also very large $\mu$) the speed of sound is given by:
\beq
c_s = {1 \over 2} \sqrt{2 \mu\over m} \; .
\eeq

In absence of two-body interactions ($g=0$) for $\rho >{1/(\pi
a_z^2)}$, i.e. for $\mu > \hbar \omega_z$, several single-particle
states of the axial harmonic well are occupied and the gas
exhibits the 2D-3D crossover. Finally, for $\rho \gg {1/(\pi
a_z^2)}$, i.e. for $\mu \gg \hbar \omega_z$, the Fermi gas becomes
3D. In this case from Eqs. (\ref{sound-c2}), (\ref{n2d-mu}) and
(\ref{s2-exact}) one finds \beq \mu = (2\pi)^{1/2}\, \hbar
\omega_z \, (a_z^2 \rho )^{1/2} \;, \label{figata-2D} \eeq \beq
c_s = \sqrt{2}\ (2\pi)^{1/4} \, a_z \omega_z \, (a_z^2 \rho)^{1/4}
\; . \label{predico} \eeq

\begin{figure}
\centering
\includegraphics[width=8cm,clip]{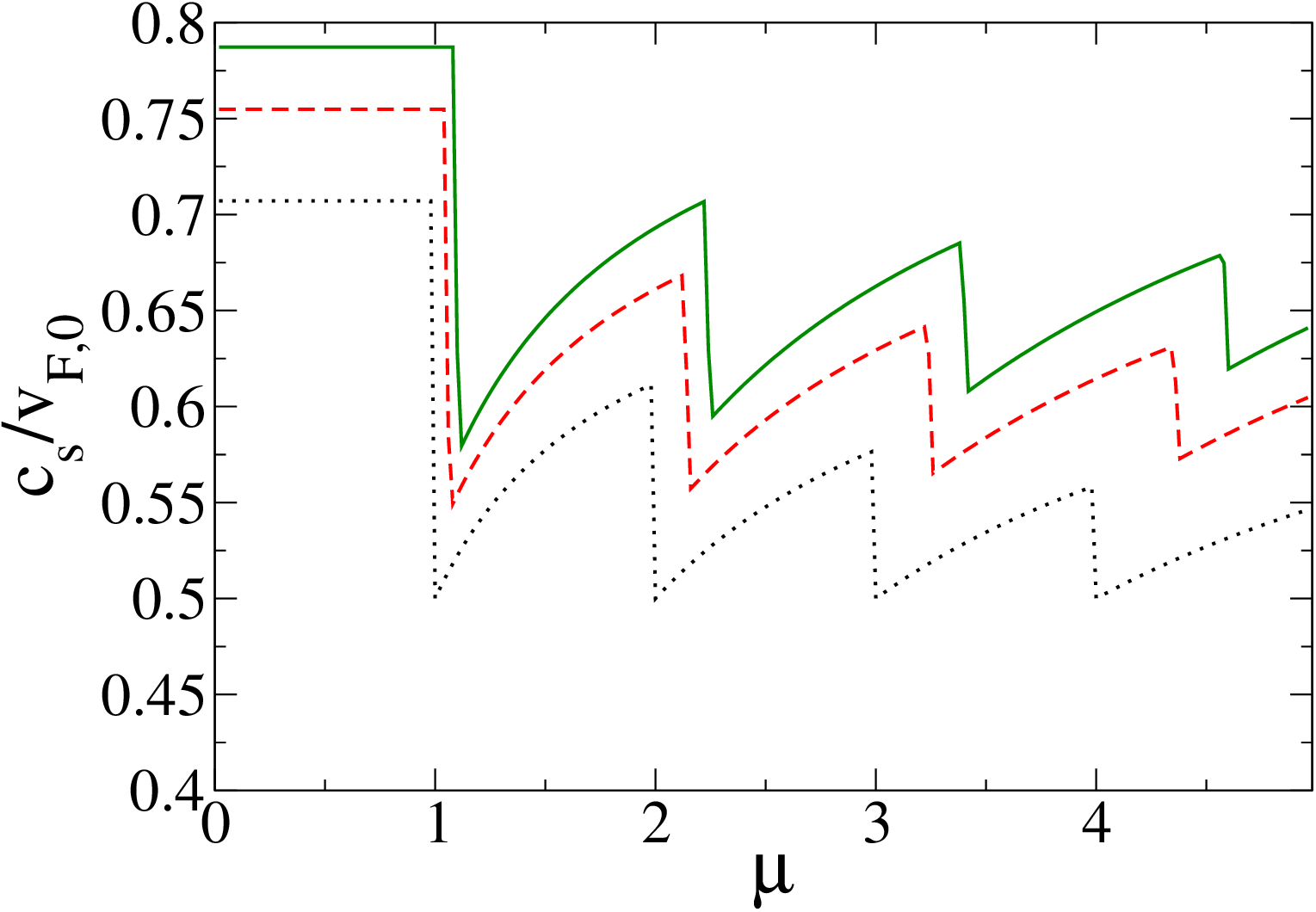}
\caption{(Color online). The ratio $c_s/v_{F,0}$ versus chemical
potential $\mu$. Chemical potential is in units of $\hbar
\omega_z$. We report behavior for $g=0$ (dotted curve),
$g=0.35$ (dashed curve), $g=0.6$ (solid curve).
The velocities $c_s$ and $v_{F,0}$ are in units of $a_z
\omega_z$, lengths in units of $a_z$, and the chemical potential in units of $\hbar \omega_z$.}
\label{Fig5}
\end{figure}

We observe that starting from the familiar Thomas-Fermi
formula for the 3D local density
$\rho(z)$ of the Fermi gas under axial harmonic confinement, given by
\beq
\rho(z) = {(2m)^{3/2}\over 3 \pi^2 \hbar^3} \left( \mu - {1\over
2} m\omega_z^2 \, z^2 \right)^{3/2} \;,
\eeq
and integrating over $z$, one obtains exactly Eq. (\ref{figata-2D}).

\section{Experimental feasibility}

We now discuss the experimental conditions to detect the zero and
first sound in a quasi-2D two-component Fermi gas. We suppose that
in the $(x,y)$ plane the system is in a square box of length
$L=20$ $\mu$m, while in the axial direction $z$ there is a
harmonic confinement characterized by a frequency of $160$ kHz,
which corresponds to a characteristic length $a_z \simeq 0.1$
$\mu$m.
We consider $^{40}$K atoms trapped in the two hyperfine states
$|f=9/2, m_f=7/2 \rangle$ and $|f=9/2, m_f=-7/2 \rangle$ with $f$
the total atomic spin and $m_f$ its axial projection. In this case
the $s$-wave scattering length reads $a_F=174$ $a_0$, where
$a_0=0.53\cdot 10^{-10}$ m is the Bohr radius \cite{regal}. For
this value of $a_F$, we can determine the value of the collective
mode frequency $\omega_c$ which discriminates between the
collisionless and the hydrodynamic regimes. In fact, the collision
time $\tau$ can be estimated as $\tau \sim 1/(\omega_z \rho
a_F^2)$ and the corresponding critical frequency is $\omega_c =
2\pi/\tau \sim \omega_z (2\pi \rho a_F^2)$. If the frequency
$\omega$ of the collective mode is much greater than $\omega_c$
(i.e. for $\omega \gg \omega_c$) the system is in the
collisionless regime; otherwise (i.e. for $\omega \ll \omega_c$)
the collective excitation is hydrodynamic-like.

One can experimentally change $\omega_c$ in the Fermi gas of
$^{40}$K atoms by varying the planar density $\rho$. For instance,
from Fig. 4 and Eq. (39) one finds that in correspondence to
$(\rho a_z^2) \pi = 1/2$ the system is strictly 2D. In this case,
the planar density is $\rho\simeq 1.6 \cdot 10^{13}$ atoms/$m^2$,
the total number of atoms is $N=\rho L^2\simeq 6.4 \cdot 10^{3}$,
and the critical frequency reads $\omega_c \simeq 1.4$ kHz. In
addition, the zero sound velocity is $c_s^{0} = v_{F,0} \simeq
0.02$ m/s, while the first sound velocity is $c_s=v_{F,0}/\sqrt{2}
\simeq 0.014$ m/s. Moreover, one finds that the scaled interaction
strength is $g = 2 a_F/a_z \simeq 0.18$. To conclude, we observe
that one can change the critical frequency $\omega_c$ also by
varying the scattering length $a_F$ via magnetic Feshbach
resonances.

\section{Conclusions}

In the present paper we have considered a dilute and ultracold interacting
disk-shaped Fermi gas confined in the axial direction by a strong harmonic
trap and uniform in the two planar directions.
In particular, we have analyzed the behavior of the velocity
of propagation of the zero and first sound modes in this gas.

In the first part of the paper, we have studied the zero sound
mode from the density-density response function within the linear
response theory and under the random phase approximation. We have
obtained the zero-sound dispersion law by calculating the poles of
the density-density response function. Then we have analyzed the
behavior of the velocity of the zero sound as a function of the
planar density for different values of the Fermi-Fermi scattering
length. We have verified that as in the strictly two dimensional
case, in correspondence of a given value of the planar density,
the zero sound velocity increases as a consequence of the
increasing of the strength of the fermion-fermion interaction. We
have studied, moreover, the behavior of the ratio between the zero
sound velocity $c_s^0$ and the planar Fermi velocity $v_{F,0}$ at
varying of the chemical potential; when, in presence of a non
vanishing fermion-fermion interaction, the chemical potential is
greater than a critical value, excited axial modes begin to be
occupied by the fermions and $c_s^0/ v_{F,0}$ grows with $\mu$. In
principle, the same kind of analysis may be carried out in
presence of a harmonic potential in the transverse radial plane
instead of the axial direction (see \cite{sala-reduce1}). In this
case, the treatment of the problem becomes much more complicated.
In fact, the cigar-shaped configuration is characterized by two
quantun mumbers related to the harmonic energetic levels in the
planar directions. However, we expect for the zero sound velocity
the same behavior as in the case of the only axial harmonic
trapping.

In the second part of the paper, from the formula which relates
the chemical potential of a disk-shaped Fermi gas to its uniform
planar density, we have calculated the collisional sound velocity
of the system. We have found that this sound velocity gives a
clear signature of the dimensional crossover of the two-component
Fermi gas. Our calculations suggest that the dimensional crossover
induces shell effects, which can be experimentally detected. In
fact, as in 1D-3D crossover \cite{sala-reduce1}, also in the 2D-3D
crossover the sound velocity exhibits jumps in correspondence of
the filling of axial harmonic modes.  We have investigated the
behavior of the ratio between the first sound velocity $c_s$ and
the planar Fermi velocity as a function of the chemical potential.
We stress that our study shows that when only the ground state of
the harmonic well is populated and when the chemical potential is
sufficiently large so many excited axial modes are occupied,  our
results reproduce the behavior of a 2D and a 3D Fermi system,
respectively.

Finally, we have discussed the possibility of observing in experiments the
zero and first sound by using gases of $^{40}$K atoms trapped in the two lowest
hyperfine states. In particular, we have estimated the sound-mode frequency
which discriminates between the collisionless and hydrodynamic
regime. We have found that these regimes require
severe geometric and thermodynamical constraints,
but they can be reached with the available experimental setups.

This work has been partially supported by Fondazione CARIPARO. The authors
thank Giovanni Modugno for useful suggestions.


\begin{thebibliography}{99}

\bibitem{book-pethick} C.J. Pethick and H. Smith,
{\it Bose-Einstein Condensation in Dilute Gases}
(Cambridge Univ. Press, Cambridge, 2002).

\bibitem{book-stringari} L.P. Pitaevskii and S. Stringari,
{\it Bose–Einstein Condensation}
(Oxford Univ. Press, Oxford, 2003).

\bibitem{demarco} B. DeMarco and D.S. Jin,
Science {\bf 285}, 1703 (1999).

\bibitem{truscott} A.G. Truscott,  K. E. Strecker,  W. I. McAlexander,
G. B. Partridge and R. G. Hulet,
Science {\bf 291}, 2570 (2001).

\bibitem{modugno} G. Modugno, G. Roati,  F. Riboli,  F. Ferlaino,
R. J. Brecha,  M. Inguscio ,
Science {\bf 297}, 2240 (2002).

\bibitem{greiner} M. Greiner, C.A. Regal, and D.S. Jin,
Nature {\bf 426}, 537 (2003).

\bibitem{joachim} S. Jochim, M. Bartenstein,  A. Altmeyer,  G. Hendl,
S. Riedl,  C. Chin,  J. Hecker Denschlag,  R. Grimm,
Science {\bf 302}, 2101 (2003).

\bibitem{gorlitz} A. Gorlitz, J. M. Vogels, A. E. Leanhardt, C. Raman,
T. L. Gustavson, J. R. Abo-Shaeer, A. P. Chikkatur, S. Gupta, S. Inouye,
T. Rosenband, and W. Ketterle
Phys. Rev. Lett. {\bf 87}, 130402 (2001).

\bibitem{schreck} F. Schreck, L. Khaykovich, K. L. Corwin, G. Ferrari,
T. Bourdel, J. Cubizolles, and C. Salomon
Phys. Rev. Lett. {\bf 87}, 080403 (2001).

\bibitem{kinoshita} T. Kinoshita, T. Wenger,
and D. S. Weiss, Science 305, 1125 (2004).

\bibitem{schneider} J. Schneider and H. Wallis,
Phys. Rev. A {\bf 57}, 1253 (1998).

\bibitem{sala0} L. Salasnich, J. Math. Phys. {\bf 41}, 8016 (2000).

\bibitem{sala1} L. Salasnich, B. Pozzi, A. Parola, L. Reatto,
J. Phys. B {\bf 33}, 3943 (2000);
L. Salasnich, L. Reatto and A. Parola,
in 'Perspectives in Theoretical Nuclear Physics VIII',
Ed. by G. Pisent et al, pp. 239-246 (World Scientific, Singapore, 2001).

\bibitem{vignolo} P. Vignolo and A. Minguzzi,
Phys. Rev. A {\bf 67}, 053601 (2003).

\bibitem{bruun} G.M. Bruun and C.W. Clark,
Phys. Rev. Lett. {\bf 83}, 5415 (1999).

\bibitem{minguzzi} A. Minguzzi, P. Vignolo, M.L. Chiofalo,
and M.P. Tosi, Phys. Rev. A {\bf 64}, 033605 (2001).

\bibitem{das} K.K. Das, Phys. Rev. Lett. {\bf 90}, 170403 (2003).

\bibitem{sala2} L. Salasnich, S.K. Adhikari, and F. Toigo
Phys. Rev. A {\bf 75}, 023616 (2007).

\bibitem{capuzzi1} P. Capuzzi, P. Vignolo, F. Federici
and M.P. Tosi, J. Phys. B: At. Mol. Opt. Phys. {\bf 37},
S91 (2004).

\bibitem{ghosh} T.K. Ghosh and K. Machida,
Phys. Rev. A {\bf 73}, 013613 (2006).

\bibitem{capuzzi2} P. Capuzzi, P. Vignolo, F. Federici
and M.P. Tosi, Phys. Rev. A {\bf 73}, 021603(R) (2006).

\bibitem{capuzzi3} P. Capuzzi, P. Vignolo, F. Federici
and M.P. Tosi, Phys. Rev. A {\bf 74}, 057601 (2006).

\bibitem{landau} L. Landau and L. Lifshitz,
{\it Course in Theoretical Physics}, Vol. 9, Statistical Physics
(Pergamon, New York, 1959).

\bibitem{pines} D. Pines and P. Noziers,
{\it The Theory of Quantum Liquids}, Vol I, Normal Fermi Liquids
(W.A. Benjamin, New York, 1966).

\bibitem{fetter} A.L. Fetter and J.D. Walecka,
{\it Quantum Theory of Many-Particle Systems}
(McGraw-Hill, Boston, 1971).

\bibitem{sala-reduce1} L. Salasnich and F. Toigo,
J. Low. Temp. Phys. {\bf 150}, 643 (2008).

\bibitem{negele} J.W. Negele and H. Orland,
{\it Quantum Many Particle Systems}
(Westview Press, Boulder, 1998).

\bibitem{vignale} G.F. Giuliani and G. Vignale,
{\it Quantum Theory of the Electron Liquid}
(Cambridge University Press, 2005).

\bibitem{lipparini} E. Lipparini,
{\it Modern Many Particle Physics} (World Scientific, 2003).

\bibitem{sala-reduce2} L. Salasnich, A. Parola, and L. Reatto,
Phys. Rev. A {\bf 69}, 045601 (2004).

\bibitem{sala-reduce3} L. Salasnich, A. Parola, and L. Reatto,
Phys. Rev. A {\bf 70}, 013606 (2004).

\bibitem{zaremba} E.Zaremba, Phys. Rev. A {\bf 57}, 518 (1998).

\bibitem{joseph}J. Joseph, B. Clancy, L. Luo, J. Kinast, A. Turlapov,
and J. E. Thomas, Phys. Rev. Lett. {\bf 98}, 170401 (2007).

\bibitem{gupta} S. Gupta, Z. Hadzibabic, J.R. Anglin, and
W. Ketterle, Phys. Rev. Lett. {\bf 92}, 100401 (2004).

\bibitem{regal} C.A. Regal and D.S. Jin, Phys. Rev. Lett. {\bf 90},
230404 (2003).


\end{thebibliography}
\end{document}